\documentclass[twocolumn, aps, pra]{revtex4-1}
\usepackage{graphicx}
\usepackage{dsfont}
\usepackage{amsfonts}
\usepackage{epstopdf}
\usepackage{amsmath,amssymb}
\usepackage{bm}
\usepackage{amsmath}
\usepackage[percent]{overpic}
\usepackage{subfig}

\begin{document}

\title{All bipartitions of arbitrary Dicke states}

\author{M. G. M. Moreno}
\author{Fernando Parisio}
\email[]{parisio@df.ufpe.br}
\affiliation{Departamento de
F\'{\i}sica, Universidade Federal de Pernambuco, Recife, Pernambuco
50670-901, Brazil}

\begin{abstract}
By exploiting the permutation symmetry of Dick states, we derive closed analytical expressions of Schmidt decompositions for {\it all} possible bipartitions of a system described by this kind of state.
This allows us to exhaustively compute the entropy of entanglement of the bipartitions and, thus, compare the their entanglement extent. 
We also address the multipartite character of Dicke states by calculating the purity of balanced bipartitions to determine the potential of multipartite entanglement (the average purity). 
In particular, we found that the entanglement of $W$ states remains constant as the number of qubits is increased.
As a final application we define a family of multipartite entanglement witnesses and compute their resistance against random and systematic imperfections. It is shown that in some
circumstances, for a fixed white noise fraction, the entanglement becomes detectable only if one {\it increases} the amount of systematic imperfection in the state.
\end{abstract}

\maketitle

\section{Introduction}

The growing interest in multipartite entangled states is partly a consequence of the rush to find, classify, and quantify useful resources to a variety of tasks in the field of quantum information. For this reason the history of several of these states is relatively recent. Outstanding examples are Greenberger-Horne-Zeilinger (GHZ) states which appeared around 1990 in the context of Bell nonlocality \cite{greenberger1990bell} and graph states \cite{graph}, particularly, cluster states which were defined in 2001 as a fundamental resource to irreversible (one-way) quantum computation \cite{briegel2001persistent,PhysRevLett.86.5188}. These states have been formally introduced in the context of quantum information and before any actual counterpart could be produced in the laboratory.

One noticeable exception is that of Dicke states. They were introduced back in 1954 in an acknowledgedly important study on the spontaneous, coherent radiation emitted by a molecular gas \cite{dicke1954coherence}. Only much later these states started to be produced in a controlled way and investigated under the perspective of quantum information \cite{PhysRevA.70.022106}. Presently, one may find several works reporting on the controlled experimental generation of this class of states, using different physical systems to encode the qubits \cite{eibl2004experimental,mikami2004generation,Wieczorek,wieczorek2009multiqubit,zhao2011robust,thiel2007generation,vanderbruggen2011spin}. 
Particularly, it is remarkable that a (single-excitation) Dicke state involving hundreds of atomic ensembles has been recently engineered in a solid-state system \cite{200}.
Some of its properties have also been studied, as for instance, nonlocal correlations \cite{YYZhao} , resistance against decoherence \cite{Guhne}, and dynamics \cite{mazets2007multiatom}.

Dicke states are symmetric with respect to qubit permutations and present recurrences, which considerably simplify the study of arbitrary partitions.
In this work we explore these symmetries to make an exhaustive analytical investigation of the entanglement associated with every possible cut. 
With this piece of information we are able to assess the capacity of Dicke states as repositories of multiparty correlations, by using the potential of multipartite entanglement (the average purity related to the reduced density matrices of balanced bipartitions) as the figure of merit. 
Finally, we obtain a family of entanglement witnesses, with which we investigate the resistance of Dicke states against preparation errors, both, random and systematic. A tradeoff appears between the amount of these two imperfections, such that, in some cases, for a fixed amount of white noise, one must increase the systematic error for the entanglement to be detectable. Since the referred witnesses are appropriate for experimental investigations, we hope that this analysis may be useful in practice.

\section{Setting one qubit apart}

Dicke states have been introduced as describing the internal degrees of freedom, ground (0) or excited (1), of the $n$ molecules composing a gas. The system is admitted to be in a volume whose characteristic dimensions are small in comparison to the wave length of the radiation (corresponding to the transitions $0\leftrightarrow 1$), but large enough for the overlap between the individual molecular wave functions to be negligible, thus, avoiding the need of symmetrization. In this regime, although the molecules are distinguishable, one cannot identify which molecule emitted or absorbed a photon. Under these conditions the state of the $n$-molecule gas with $k$ excited molecules is 
\begin{eqnarray}
\label{def}
|D^{(k)}_n\rangle= \left.{n}\choose{k}\right.^{-\frac{1}{2}}\sum_{q=1}^{{n}\choose{k}}\hat P_q^{(n)}|\underbrace{0...0}_{n-k}\underbrace{1...1}_{k}\rangle,
\end{eqnarray}
which is the general form of a Dicke state \cite{dicke1954coherence}. The operator $\hat P_q^{(n)}$ performs the $q$th nontrivial permutation on the $n$ entries of the ket.

\begin{figure}[h]
	\includegraphics[scale=0.34]{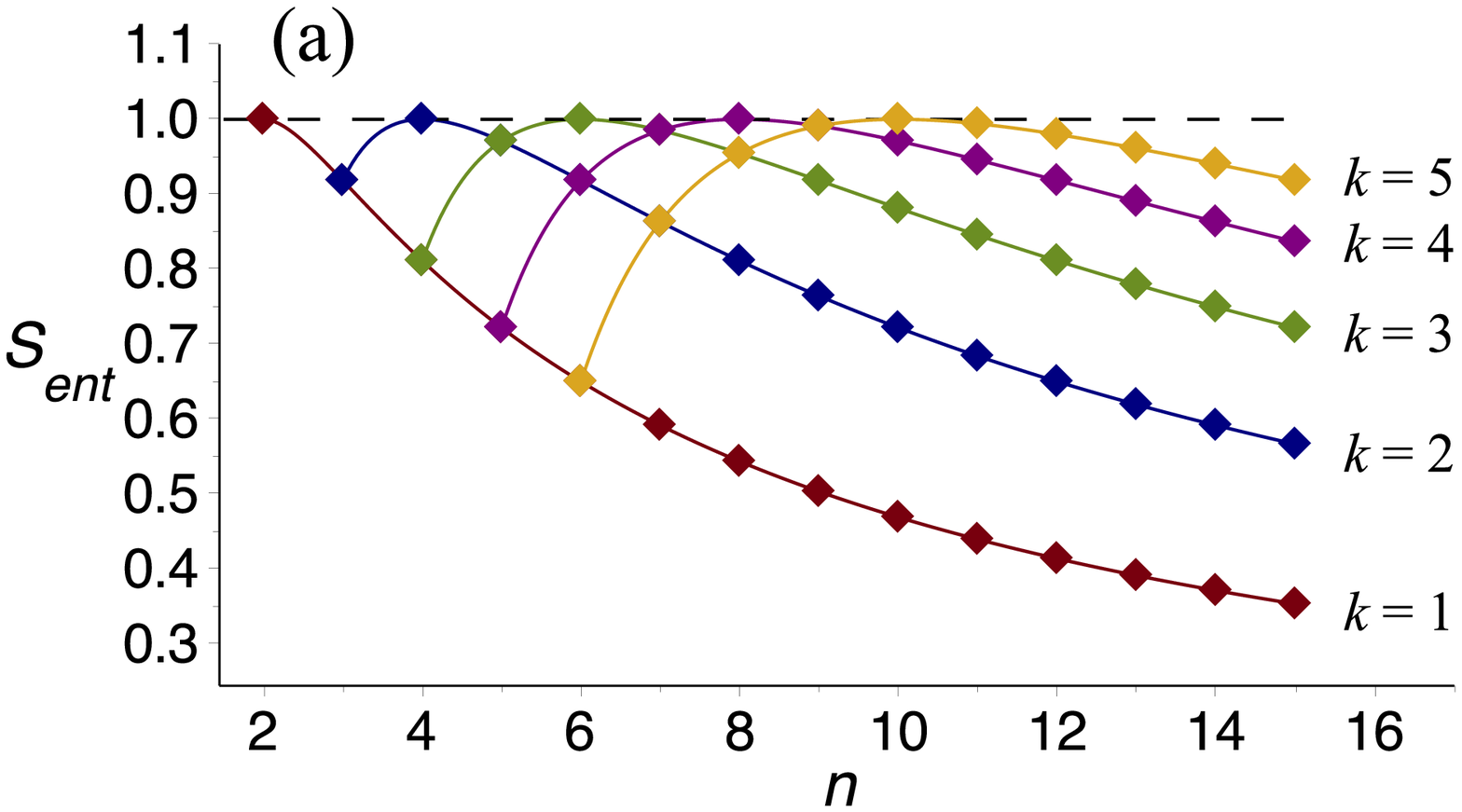}
	\includegraphics[scale=0.34]{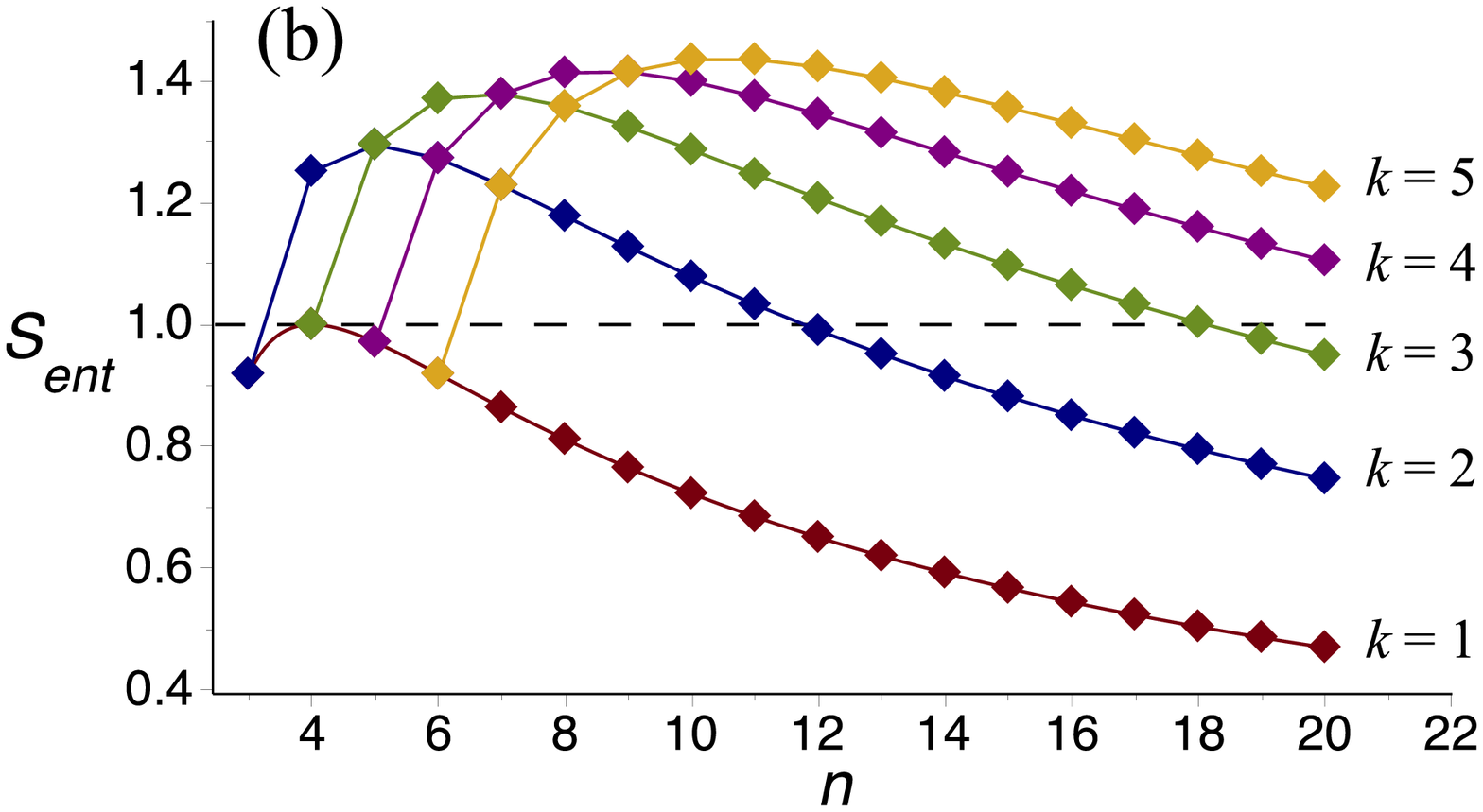}
	\includegraphics[scale=0.34]{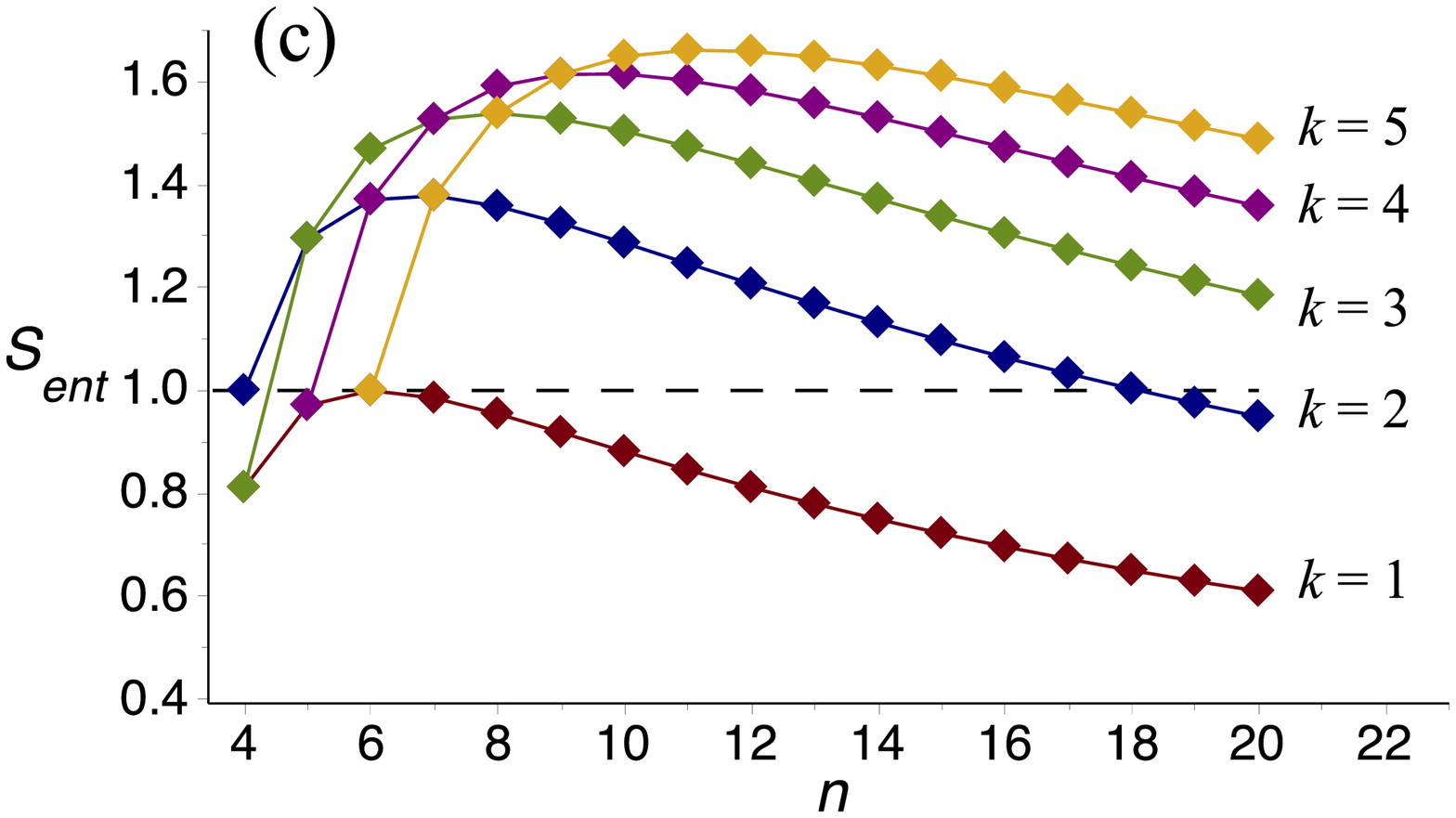}	
	\caption{(color online) In panel (a) we display the entanglement of a single qubit belonging to a Dicke state with the $(n-1)$ remaining qubits, as a function of $n$, for selected values of $k$. The plots in (b) and (c) represent the same quantities, but with 2 and 3 qubits set apart, respectively. The horizontal lines represent the maximal entanglement that a single qubit may have with any other system.}
	\label{fig1}
\end{figure}

The starting point of our study is the observation that definition (\ref{def}) can be rewritten as:
\begin{eqnarray}
\nonumber
|D^{(k)}_n\rangle  =  \left. n\choose k\right.^{-\frac{1}{2}}|0\rangle 
\otimes\sum_{q=0}^{n-1\choose k}\hat P_q^{(n-1)}|\underbrace{0...0}_{n-k-1}\underbrace{1...1}_k\rangle+\\
\nonumber
 \left. n\choose k\right.^{-\frac{1}{2}}|1\rangle
  \otimes \sum_{q=0}^{n-1\choose k-1}\hat P_q^{(n-1)}|\underbrace{0...0}_{n-k}\underbrace{1...1}_{k-1}\rangle,
\end{eqnarray}
or simply
\begin{equation}
\label{first}
|D^{(k)}_n\rangle  =  \left(\frac{n-k}{n}\right)^{\frac{1}{2}}|0\rangle |D^{(k)}_{n-1}\rangle+
\left(\frac{k}{n}\right)^{\frac{1}{2}}|1\rangle |D^{(k-1)}_{n-1}\rangle,
\end{equation}
where we dropped the tensor product symbol $\otimes$ (we will do so hereafter). First note that, due to the symmetry of the Dicke state, there is no need to specify which qubit is being singled out from the other $(n-1)$ qubits in the previous equations. All partitions of the system into 1 qubit and $(n-1)$ qubits, which we denote by $(1|n-1)$, are equivalently described by the state (\ref{first}). In addition, Eq. (\ref{first}) is a Schmidt decomposition for the referred bipartition. This allows us to use the well-settled theoretical apparatus concerning the entanglement of pure states of bipartite systems (the fact that the two Hilbert spaces have different dimensions is immaterial). One can immediately write the entropy of entanglement, $S$, bearing the partition $(1|n-1)$:
\begin{equation}
\nonumber
S(n,k,1)  =  -\left(\frac{n-k}{n}\right)\log\left(\frac{n-k}{n}\right)-\left(\frac{k}{n}\right)\log\left(\frac{k}{n}\right),
\end{equation}
where the last argument in $S$ indicates that one of the partitions has a single qubit.
Given the total ``size'' of the system $n$, $S$ attains its maximum value at $k=n/2$ [$k=(n\pm 1)/2$] for $n$ even (odd). For $n$ even, $S(n,n/2,1)=1$ ebit, so that, the the qubit held by Alice is maximally entangled with Bob's part. For $n$ odd, $S(n,(n\pm 1)/2,1)=1-O(1/n)$. On the other hand, as the number of qubits grows and the number of excitations $k$ remains fixed, the amount of entanglement shared between Alice and Bob decreases and approaches zero as $n\rightarrow\infty$. A plot of the entropy of entanglement $S(n,k,1)$ as a function of $n$, for different values of $k$, is shown in figure \ref{fig1}(a).
Notice also that the first points of a given $k$-constant curve superimpose points of other curves referring to smaler values of $k$. This is a consequence of the fact that Dicke states $|D^{(k)}_n\rangle$ and $|D^{(n-k)}_n\rangle$ are related trough local operations and classical communication (LOCC).
This drop in the entanglement as $n$ grows and $k$ remains fixed was already observed in reference \cite{moreno2016optimal}.
\section{All Bipartitions}
One can use Eq. (\ref{first}) self-consistently to expand $|D^{(k)}_{n-1}\rangle$ and $|D^{(k-1)}_{n-1}\rangle$ in terms of Dicke states with $k$, $k-1$, and $k-2$ excitations and so forth. By repeating this procedure $j$ times we obtain:
\begin{eqnarray}
\label{gen}
\nonumber
|D^{(k)}_n\rangle  &=&\sqrt{n!}\left[\left. n\choose k\right. \left. n\choose j\right.\right]^{-1/2}\\
& &\sum_{q=q'}^{q''}  \frac{|D^{(q)}_j\rangle|D^{(k-q)}_{n-j}\rangle}{\sqrt{q!(j-q)!(k-q)!(n-k-j+q)!}},
\end{eqnarray}
with $q'=\max(0,j+k-n)$  and $q''=\min(j,k)$ ($q''>q'$). The validity of this expression can be proved, for any finite $n$, $j$, and $k$, via the principle of finite induction (see the Appendix). Note that Eq. (\ref{gen}) is a Schmidt decomposition for all bipartitions $(j|n-j)$. We, therefore, can immediately write the entropy of entanglement corresponding to any bipartition as
\begin{eqnarray}
\label{Sj}
\nonumber
S(n,k,j)= -\sum_{q=q'}^{q''}  \frac{(n-j)!(n-k)!q!}{(n-k-j+q)!n!}\left. j\choose q\right. \left. k\choose q\right.\\
 \log\left[\frac{(n-j)!(n-k)!q!}{(n-k-j+q)!n!}\left. j\choose q\right. \left. k\choose q\right. \right],
\end{eqnarray}
where the complete symmetry between $j$ and $k$ is evident (also in the definitions of $q'$ and $q''$). We stress that, in general, $S(n,k,j)$ is different from $S(n,k^{\prime},j)$, except in the mentioned case where $k^{\prime}=n-k$, and these quantities cannot be led to coincide via LOCC. {\it Therefore, Dicke states with different excitations $k$ and $k'$ are LOCC inequivalent (for $k'\ne n-k$)}. In Figs. \ref{fig1}(b) and \ref{fig1}(c) we plot $S(n,k,j)$ as functions of $n$, for selected values of $k$ and $j=2$ and $j=3$, respectively.

\begin{figure}[!ht] 
    \subfloat{%
      \begin{overpic}[width=0.34\textwidth]{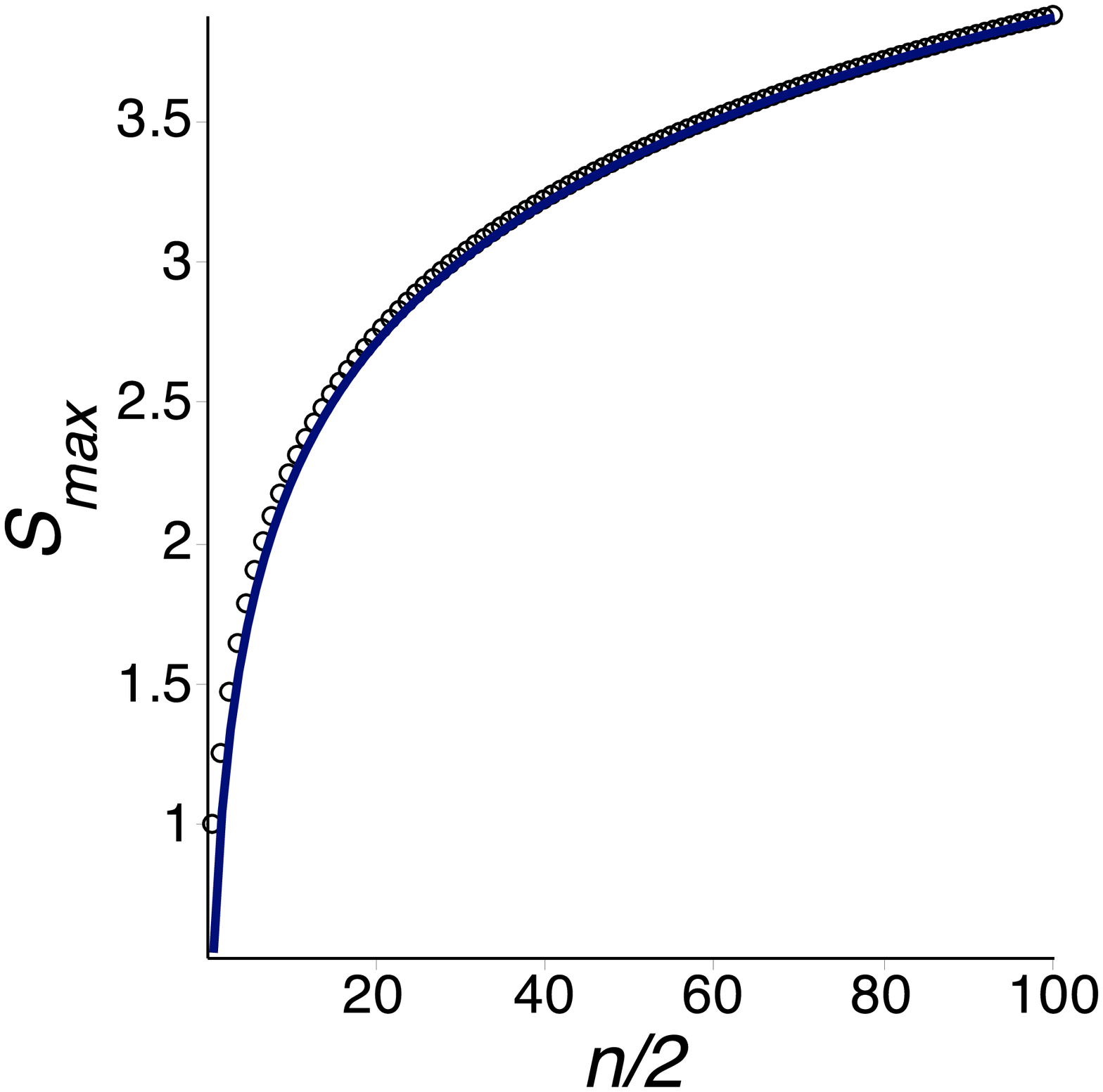}
        \put(42,19){\includegraphics[width=55\unitlength]{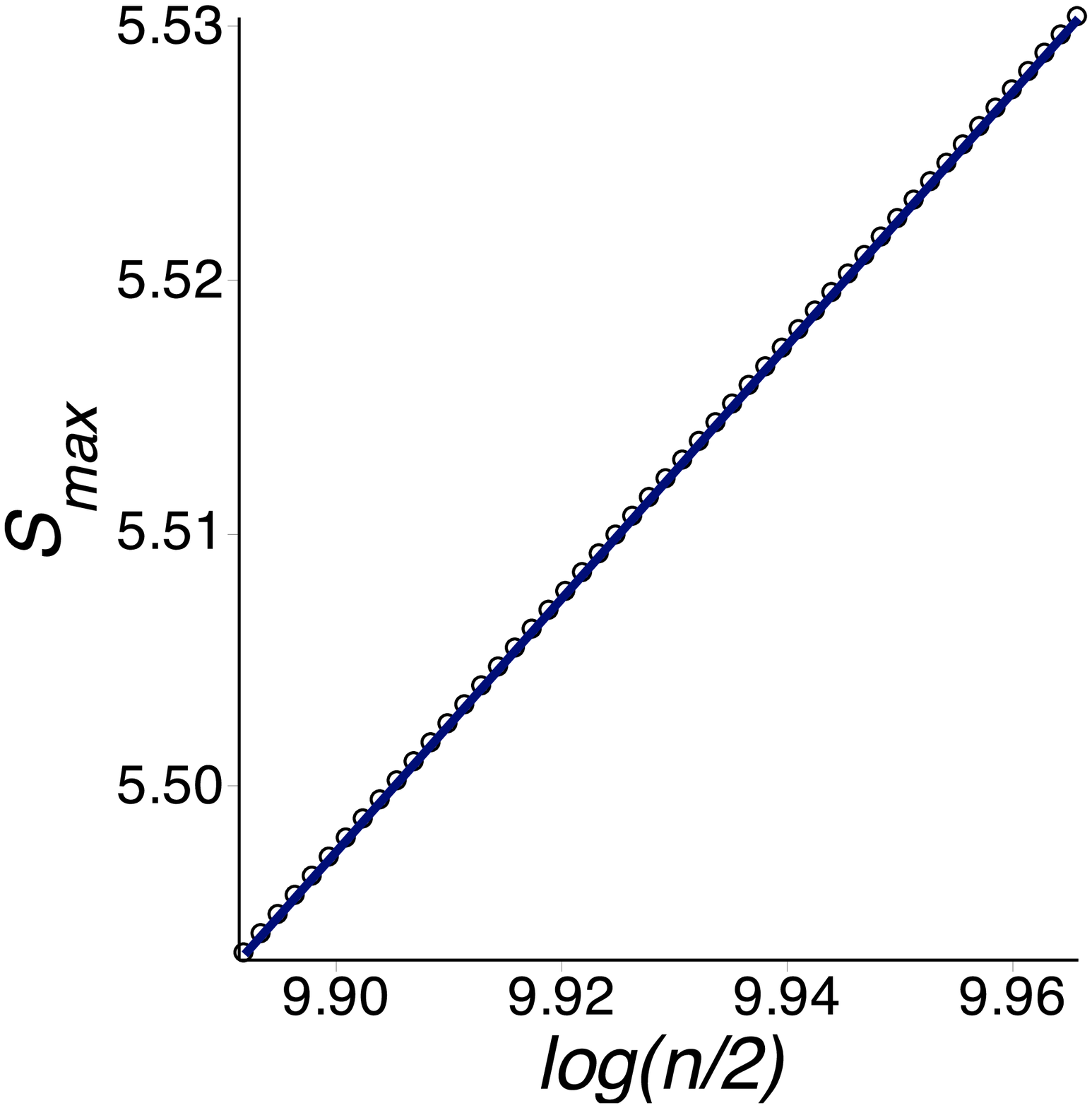}}  
      \end{overpic}
    }
    \caption{(color online) Uper bound for the entropy of entanglement associated with bipartitions of quantum systems described by Dicke states as a function of $j=n/2$ ($n$ even).
The inset shows a plot of $S_{max}$ against $\log(n/2)$ for $1900\le n \le 2000$. In both panels the continuous line represents the fitting $S=0.50 \log(n/2)+ 0.5475$.}
\label{figure2}
\end{figure}

Although the previous expression is not very transparent regarding the dependence of $S$ on the total energy ($\sim k$), the total size of the system ($\sim n$), and the proportion between the bipartitions ($j$), one property stands out: No matter the bipartition under consideration and the value of $k$, $S$ increases slowly with the size of the system. There is a trivial, generic upper bound since two quantum systems with dimensions $d_1$ and $d_2$ can be seen as two {\it qudits}, whose maximal entanglement is given by
$\log$[min$\{d_1,d_2\}]$, where, in the present case, $d_1=2^j$ and $d_2=2^{n-j}$, so that, in general, 
\begin{equation}
S\le \frac{\log 2}{2} n,
\end{equation}
for any bipartition.
In the specific case we have at hand, given the equivalence $|D^{(k)}_n\rangle\overset{LOCC}{\longleftrightarrow}|D^{(n-k)}_n\rangle$ the function $S(n,k,j)$ presents a maximum at $k=n/2$, and, thus, due to the $j$-$k$ symmetry, at $j=n/2$, for $n$ even (the conclusions are unchanged for $n$ odd). We, therefore, have $S(n,k,j)\le S(2j,j,j)\equiv S_{max}(n=2j)$. This maximal entropy can be written as 
\begin{equation}
\label{Smax}
S_{max}= \log \left. n\choose n/2\right.-\frac{2}{\left. n\choose n/2\right.}\sum_{q=0}^{n/2} \left. n/2\choose q\right.^2
\log\left. n/2\choose q\right. ,
\end{equation}
for a system with $n=2j$ qubits. A plot of $S_{max}$ as a function of $j=n/2$ is shown in figure \ref{figure2}. The data display a logarothmic asymptotic behaviour:
\begin{equation}
\label{Smaxasymp}
S_{max} \longrightarrow  \frac{1}{2}\log (n/2),\; {\rm as} \; n \rightarrow \infty.
\end{equation}
The continuous curve is a fitting with $S=0.50 \log(n/2)+ 0.5475$ and the inset shows a plot of $S_{max}$ against $\log(n/2)$ for $1900\le n \le 2000$, where the continuous line corresponds to the same fitting.
So, the entropy of entanglement related to arbitrary bipartitions of general Dicke states is not additive with respect to the number of qubits, as is a thermodynamic entropy. 
In the special case of Dicke states this comparison is not out of place. The point, specific to Dicke states, $|D^{(k)}_n\rangle$ is a balanced superposition of all possible classical microstates compatible with the macroscopic constraint that defines the microcanonical ensemble (fixed energy and particle number). In fact, a completely decohered Dicke state exactly describes a classical system of $n$ particles with total energy $E=\epsilon k$, where $\epsilon$ is the energy difference between the individual excited and ground states. Explicitly, under full decoherence we obtain
\begin{equation}
\label{dec}
|D^{(k)}_n\rangle \overset{full \;dec.}{\longrightarrow} \rho^{(k)}_n=\frac{1}{\left. n\choose k\right.}\mathds{1}_{E},
\end{equation}
where $\mathds{1}_{E}$ denotes the identity operator in the space of fixed energy ($E=k \epsilon$). The state in the right-hand side obeys the postulate of equal probabilities {\it a priori} and corresponds to a genuine microcanonical equilibrium state, thus having a classically well defined internal energy and Boltzmann entropy (${\cal S}$). It is straightforward to show that the von Neumann entropy of $\rho^{(k)}_n$ becomes the Boltzmann entropy. Using a system of units such that $k_B=1$  and taking the logarithms in the binary basis, the thermodynamic entropy associated with any bipartition of a classical system described by the state in the right-hand side of (\ref{dec}) 
grows linearly with the size of the total system, $n$, as it should. Therefore, $S_{max}/{\cal S} \longrightarrow 0$ as $n \rightarrow \infty$. Bipartitions  associated to the microcanonical constraint are more relevant as entropic repositories in the classical case.
\section{mutltipartite entanglement}
\subsection{The potential of multipartite entanglement}
So far we have been concerned with the entanglement related to arbitrary bipartitions, however, with the previous results it is possible to analyse some aspects of the multipartite entanglement of $|D^{(k)}_n\rangle$ itself. In 2008 Facchi et al \cite{facchi}, based on an early work by Scott \cite{scott}, proposed a way to assess the multipartite entanglement of pure states, which becomes extremely simple in the present case. The relevant figure of merit is inspired in the fact that, for bipartite systems described by pure states, the larger the entanglement, the more mixed is the reduced density matrix of either of the two subsystems. It is reasonable to think that, for a multipartite system, if all possible {\it balanced} bipartitions lead to maximal mixedness in the reduced subsystems, then, the whole state has a large entanglement. By a balanced bipartition the authors of \cite{facchi} mean any cut $(n/2|n/2)$ $[(n/2+1/2|n/2-1/2)]$ for $n$ even (odd). 
Let us denote the purity of a particular balanced bipartition $K$ of a multiparty state $|\Psi\rangle$ as $\varPi_K$
The {\it potential of multipartite entanglement}  is defined as the average purity of all balanced bipartitions:
\begin{equation}
\varPi_{\rm ME}(|\Psi\rangle) =\frac{1}{\left. n\choose n/2\right.}\sum_K\varPi_K.
\end{equation}
In this context the maximally entangled state is defined in \cite{facchi} as a minimizer of $\varPi_{\rm ME}$. 
Of course, this definition only captures certain aspects of multipartite entanglement, and the term ``maximally entangled'' has no absolute meaning.
For two and three qubits, the minimizers are Bell and Greenberger-Horne-Zeilinger (GHZ) states, respectively. These are examples of {\it perfect}
minimizers, since the lower bound attains the minimal algebraic value of $2^{-n/2}$. This is also possible for five and six qubits, but in all other cases (4 qubits or more than
6 qubits), it is a mathematical impossibility to attain the algebraic minimum in all balanced bipartitions (It is curious that the case involving 7 qubits was closed a few months ago \cite{huber}). The minimization of the purity associated with some bipartitions precludes full minimization in some other bipartition(s), i. e., frustration takes place.

From our previous discussion on Dicke states it is immediate that all balanced bipartitions are completely equivalent and, consequently, all $\varPi_K$ are equal.
So, we just have to calculate the purity of one (any) balanced bipartition to get $\varPi_{\rm ME}(|D^{(k)}_n\rangle)$.
For $n$ even, the reduced density operator of a balanced bipartition is given by
\begin{displaymath}
\sigma_{R}=\frac{1}{\left. n\choose k\right.}\sum_{q=0}^{k} \left. n/2\choose k-q\right.\left. n/2\choose q\right.|D^{(q)}_{n/2}\rangle \langle D^{(q)}_{n/2}|,
\end{displaymath}
where $1 \le k \le n/2$. Therefore, the associated purity, $\varPi={\rm Tr}\sigma_{R}^2$, is exactly $\varPi_{\rm ME}(|D^{(k)}_n\rangle)$ and reads
\begin{equation}
\varPi_{\rm ME}(|D^{(k)}_n\rangle)=\left. n\choose k\right.^{-2}\sum_{q=0}^{k}\left[ \left. n/2\choose k-q\right.\left. n/2\choose q\right. \right]^2,
\end{equation}
from which it is clear that $\varPi_{\rm ME}(|D^{(k)}_n\rangle)\le\varPi_{\rm ME}(|D^{(n/2)}_n\rangle)$, the equality corresponding to $k=n/2$. The extremal situations are $k=1$ and $k=n/2$. From the previous equation it is easy to show that $\varPi_{ME}$ of a $n$-qubit Dicke state with a single excitation, a $W$ state, does not depend on $n$:
\begin{equation}
\varPi_{\rm ME}(|W_n\rangle)=\frac{1}{2}\;\;\; \forall \;\;n.
\end{equation}
Therefore, according to this quantifier, the entanglement of a $W$ state does not increase with the number of qubits. In the other extreme, when the number
of excitations corresponds to half of the number of particles, we have the Dicke state with minimal potential of multipartite entanglement (recall that maximal multipartite entanglement is associated with the minimization of $\varPi_{ME}$). In this case
\begin{equation}
\varPi_{\rm ME}(|D^{(n/2)}_n\rangle)=\left. n\choose n/2\right.^{-2}\sum_{q=0}^{n/2}\left. n/2\choose q\right.^4,
\end{equation}
In the limit of large $n$, one can use the asymptotic result $\sum_{q=0}^{\ell}\left. \ell \choose q\right.^{\nu} \sim (2/\pi \ell)^{(\nu-1)/2}\,\nu^{-1/2}\,2^{\nu \ell}$. This result, together with the asymptotic limit of the combinatorial symbol, lead to
\begin{equation}
\varPi_{\rm ME}(|D^{(n/2)}_n\rangle)\sim \frac{2}{\sqrt{\pi}} \,n^{-1/2}, \; n \rightarrow \infty.
\end{equation}
Therefore, while $\varPi_{\rm ME}$ of Dicke states with a single excitation remain constant as $n$ grows, 
it asymptotically decreases as $n^{-1/2}$ for Dicke states with $k=n/2$. Again, for $n$ odd the results are qualitatively the same.
These values are much larger than numerically calculated minimizers, which, even with frustration, attain values close to the
algebraic minimum of $2^{-n/2}$.
\subsection{Multipartite entanglement witnesses}
Here we Bourennane \textit{et al.}  \cite{bourennane2004experimental} provided a scheme for the construction of multipartite entanglement witnesses \cite{horodecki1996separability,horodecki2009quantum}. Given an entangled state $|\Psi\rangle$ the witness they build would be able to detect entanglement of $|\Psi\rangle$ and states close to it. The witness $\mathcal{W}$ reads:
\begin{eqnarray}
{\cal W}=\alpha\mathds{1} - |\Psi\rangle\langle\Psi|,
\end{eqnarray}
where $\alpha$ is the square modulus of the highest Schmidt coefficient over all possible bipartitions. Whenever we find a state $ |\Psi\rangle$ for which $\langle {\cal W} \rangle<0$, then, multipartite entanglement is certified.

Since we have explicit Schmidt decompositions for all possible bipartitions of a $n$-partite system the representation of such witness is immediate. Considering that the partition that maximizes the Schmidt coefficient must be the same that minimizes the entropy, $\alpha$ corresponds to $j=1$. Given the LOCC equivalence between Dicke states of $n$ qubits with $k$ and $n-k$ excitations, we will only consider $1\leq k\leq n/2$. Therefore, the explicit form of the family of witnesses we consider here is
\begin{eqnarray}
{\cal W}^{(k)}_n=\left(\frac{n-k}{n}\right)\mathds{1} - |D^{(k)}_n\rangle\langle D^{(k)}_n|,
\label{wit}
\end{eqnarray}
A direct application of those witnesses is to test the entanglement resistance of Dicke states in the presence of noise. Consider a state $\rho_k=p\frac{1}{2^n}\mathds{1} + (1-p)|D_n^{(k)}\rangle\langle D_n^{(k)}|$, the question is how large can $p$ be so that we can still witness entanglement with $\mathcal{W}_k$. One can easily show that the maximum value of white-noise admixture:
\begin{equation}
\label{p}
p_{max}=
\frac{k}{n\left(1-2^{-n}\right)}.
\end{equation}
This improves on previous results by T\'oth specifically for $k=1$ and $k=n/2$ \cite{toth2007detection}, but is valid for any $k$ with $1 \le k \le [n/2]$. For $[n/2]<k\le n$, in the previous equation, we must take $ k \rightarrow n-k$. 

Another possibility is to verify the influence of a systematic asymmetry in a Dicke state. For instance, let us consider the state 
\begin{eqnarray}
\nonumber
|\phi_n^{(k)}\rangle=(1-a^2)^{\frac{1}{2}}|D_n^{(k)}\rangle+a|\underbrace{0...0}_{n-k}\underbrace{1...1}_{k}\rangle,
\end{eqnarray}
where $0 \le a \le 1$ sets the amount of asymmetry in the state. Note that, we consider an energy-preserving asymmetry, since the second term in the previous expression is also related to $k$ excitations \cite{comm}. For $a=0$ we simply get $|\phi_n^{(k)}\rangle=|D_n^{(k)}\rangle$, while for $a=1$ 
the state is separable.
Using definition (\ref{wit}), we can easily show that the expectation value of witness (\ref{wit}) is $$\langle \phi_n^{(k)}|{\cal W}^{(k)}_n|\phi_n^{(k)}\rangle=\frac{n-k}{n}-\left[ \sqrt{1-a^2}+a\left. n\choose k\right.^{-1/2}\right]^2,$$
from which it is easy to check that $|D_n^{(n/2)}\rangle$ has a more robust entanglement as $a$ grows in comparison to other Dicke state, especially to $|W_n\rangle$.
This is, in fact, expected in the light of the results in the previous sections.

\begin{figure}[h]
	\includegraphics[scale=0.215]{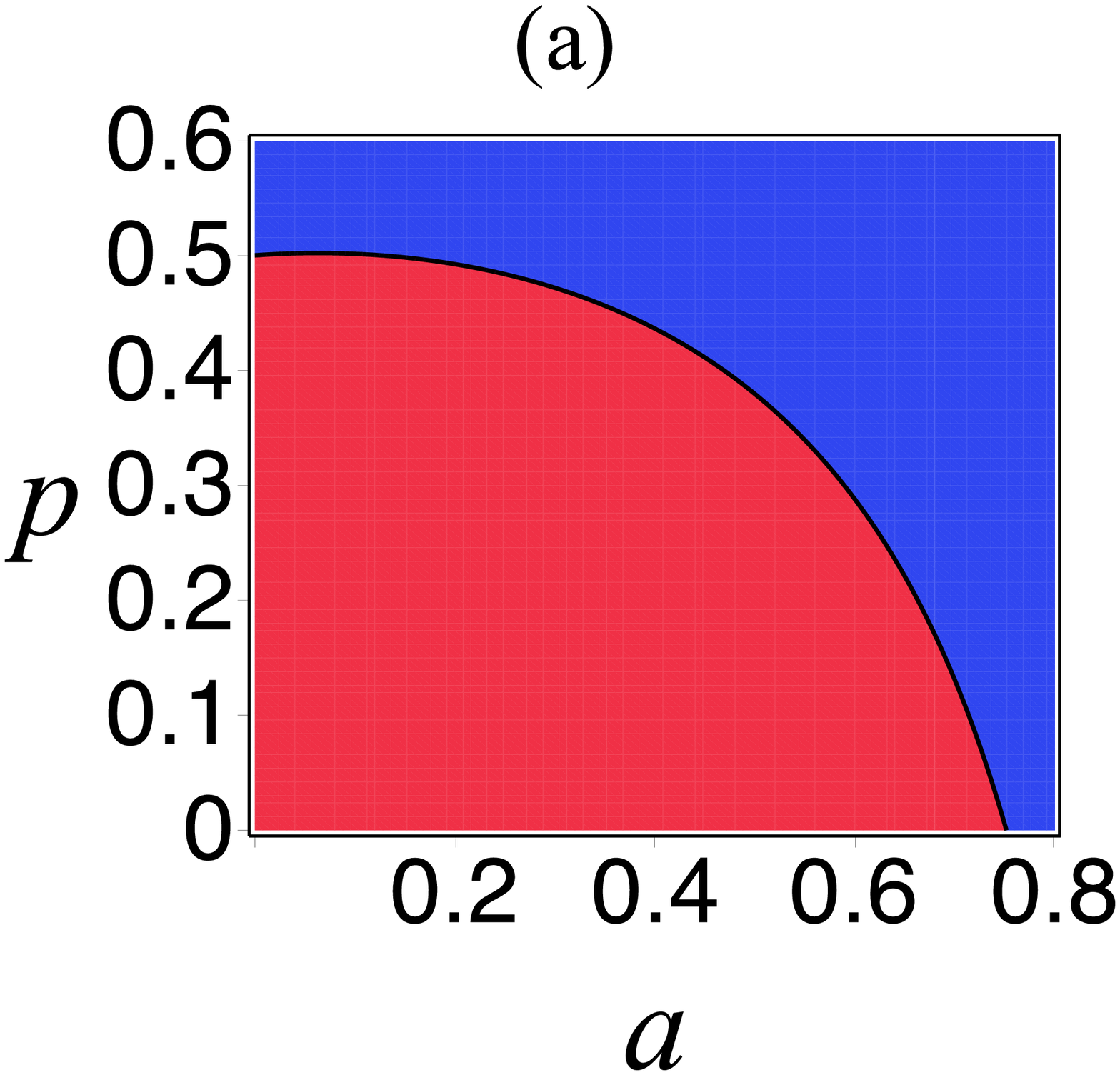}
	\includegraphics[scale=0.215]{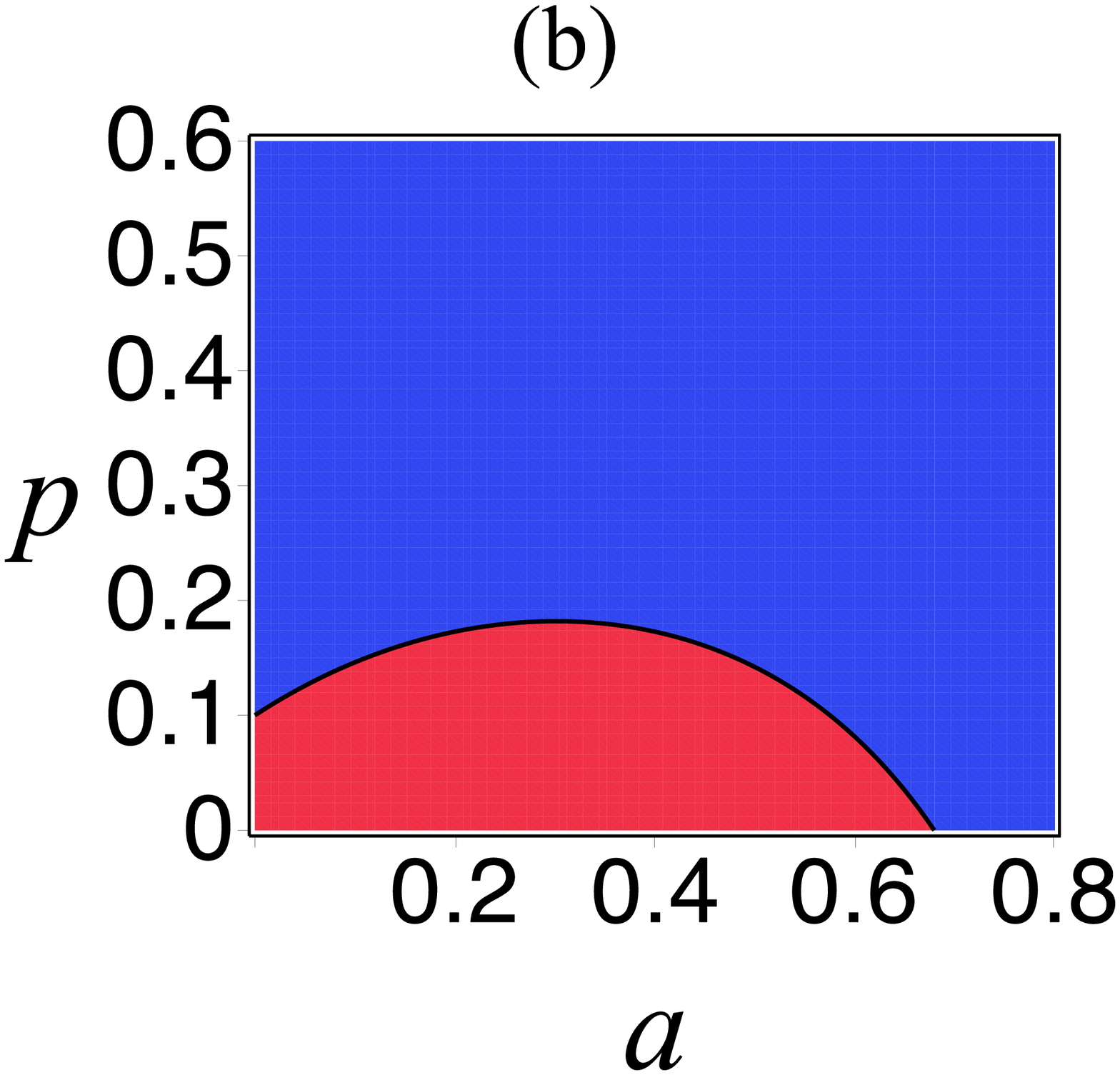}
	\caption{(color online) Regions in the plane of parameters $a$-$p$ for which $\langle {\cal W}^{(k)}_n\rangle<0$ (below the continuous curve) and $\langle {\cal W}^{(k)}_n\rangle>0$ (above the continuous curve). Panel (a) refers to the state $|D_{10}^{(5)}\rangle$, and panel (b) to the state $|D_{10}^{(1)}\rangle=|W_{10}\rangle$.}
	\label{fig3}
\end{figure}

A more interesting situation appears when we consider that the states of interest present both, some systematic and random imperfections simultaneously. This is clearly a situation of practical interest. The imperfect state reads
\begin{equation}
\label{imp}
\varrho_k= (1-p)|\phi_n^{(k)}\rangle\langle \phi_n^{(k)}|+p\frac{\mathds{1}}{2^n} 
\end{equation}
and elementary calculations to evaluate Tr$(\varrho_k {\cal W}^{(k)}_n)=\langle {\cal W}^{(k)}_n\rangle$, lead to
\begin{equation}
\frac{n-k}{n}-\frac{p}{2^n}-(1-p)\left[ \sqrt{1-a^2}+a\left. n\choose k\right.^{-1/2}\right]^2.
\end{equation}
In fig. \ref{fig3} we show the regions in the $a$-$p$ plane in which multipartite entanglement can be certified (red) and the regions 
for which $\langle {\cal W}^{(k)}_n\rangle>0$ (blue). Panel (a) refers to the state $|D_{10}^{(5)}\rangle$, and panel (b) to the state $|D_{10}^{(1)}\rangle=|W_{10}\rangle$.
As expected the former state is more resilient than the latter, also with respect to the witness defined by Eq. (\ref{wit}).

The behaviour of the separatrix $\langle {\cal W}^{(k)}_n\rangle=0$ (continuous curve) is not monotonous in the variable $p$ as $a$ varies. This is particularly evident for the imperfect $W$ state, fig. \ref{fig3}(b), for which, in the absence of any systematic error ($a=0$), the maximum amount of white noise tolerated is $p_{max}\approx 0.10$. However, for $a\approx 0.30$ the maximum noise fraction becomes $p_{max}\approx 0.18$. So, if we have some basal white noise fraction $p$, not easily removable, then if $0.10<p<0.18$, we may, purposefully, increase the amount of asymmetry of state (\ref{imp}) in order to make its entanglement detectable by the witness (\ref{wit}).  This effect is also present for the state with $k=5$, fig. \ref{fig3}(a), but it is much less pronounced: while for $a=0$ we have $p_{max} \approx 0.500$, for $a\approx 0.063$ we get $p_{max} \approx 0.502$.

One can understand this effect by rewriting (\ref{imp}) as $\varrho_k=(1-p)(1-a^2)|D_n^{(k)}\rangle\langle D_n^{(k)}| + [(1-p)a^2+p]\rho_{color}$, where $\rho_{color}$ can be easily determined through (\ref{imp}). So, there is a compromise: On the one hand, by increasing $a$ we decrease the extent of $|D_n^{(k)}\rangle$ in the mixture, and, on the other hand, we weaken the effect of white noise, by giving it some color. Under this perspective, it is not surprising that, given some amount of white noise, the optimal value of $a$ is nonzero. 

Finally, it is well known that by tracing over one of the quibits in a $W$ state, the remaining system is still entangled. This result is generalized here, for, it is easy to prove that by tracing $|D_n^{(k)}\rangle$ over a single qubit, the witness $ {\cal W}^{(k)}_{n-1}$ is still able to detect the remaining entanglement.

\section{Summary and Conclusion}
Dicke states are completely symmetric with respect to permutation of particles, which leads to recurrences between states involving different number of particles and excitations.
By using these recurrences, we provide a closed form for Schmidt decompositions of all bipartitions of a $n$-qubit system in an arbitrary Dicke state. We found an upper bound for the entropy of entanglement related to these bipartitions and showed that $S$(arbitrary bipartition of $|D_n^{(k)}\rangle)<(1/2)\log (n/2)$ as $n\rightarrow \infty$. After full decoherence, $|D_n^{(k)}\rangle$ becomes a classical microcanonical state in the thermodynamical limit
of $n, k \rightarrow \infty$ with $k/n$ finite. In this case, we can compare the amount of entropy before and after decoherence takes place: Bipartitions are more entropic in the classical case than in the quantum domain.

It is possible to detect and partially quantify multipartite entanglement solely on the basis of an exhaustive knowledge on the bipartitions of the system of interest.
We, therefore, used our characterization to calculate the potential of multipartite entanglement of Dicke states. While, according to this measure, the
entanglement of a $W$ state doesn't depend on the number of qubits, the potential of a balanced state, with $k=[n/2]$, decreases as $n^{-1/2}$ for large $n$.
This value is away below the minimum algebraic value, which goes as $2^{-n/2}$. 
We also defined a family of entanglement witnesses, whose usefulness has been demonstrated with simple examples. It is hoped that these results can be used
in the laboratory since the witness defined in \cite{bourennane2004experimental}, which in our work assumes the form (\ref{wit}), is particularly amenable to experimental realizations.
\section{Appendix}
To demonstrate relation (\ref{gen}) we use the principle of finite induction. Thus, we start proving its validity for $j=1$, then we show that if it is valid for an arbitrary value of $j$ it holds for $j+1$.

First, let us show that it is valid for $j=1$. We write $|D_n^{(k)}\rangle$ as
\begin{eqnarray}
\nonumber
 \sum_{q=0}^{1}\left[\frac{(n-1)!k!(n-k)!}{n!(k-q)!(n-k-1+q)!}\left.j\choose q\right.\right]^{\frac{1}{2}}
|D^{(q)}_1\rangle|D^{(k-q)}_{n-1}\rangle\\
\nonumber
 = \left[\frac{(n-1)!k!(n-k)!}{n!k!(n-k-1)!}\right]^{\frac{1}{2}}|D_1^{(0)}\rangle|D_{n-1}^{(k)}\rangle+\\
\nonumber
 \left[\frac{(n-1)!k!(n-k)!}{n!(k-1)!(n-k)!}\right]^{\frac{1}{2}}|D_1^{(1)}\rangle|D_{n-1}^{(k-1)}\rangle\\
\nonumber
 =  \left(\frac{n-k}{n}\right)^{\frac{1}{2}}|0\rangle|D_{n-1}^{(k)}\rangle+\left(\frac{k}{n}\right)^{\frac{1}{2}}|1\rangle|D_{n-1}^{(k-1)}\rangle,
\end{eqnarray}
which agrees with relation (\ref{first}) presented in the body of the text.
Now we are going to test if assuming its validity for some arbitrary $j$ it is also valid for $j+1$. Rigorously, we have to consider four different situations: $j\leq k$ and $0>j-n+k$, $j> k$ and $0>j-n+k$, $j\leq k$ and $0\leq j-n+k$, and  $j> k$ and $0\leq j-n+k$. Here we address the case  $j\leq k$ and $0>j-n+k$, similar steps can be performed in all other situations.
Inserting relation (\ref{first}) into (\ref{gen}) we get:
\begin{eqnarray}
\nonumber
|D_n^{(k)}\rangle & = & \sum_{q=0}^{j}\left[\frac{(n-j)!k!(n-k)!}{n!(k-q)!(n-k-j+q)!}\left.j\choose q\right.\right]^{\frac{1}{2}}\\
\nonumber
& & \left(\frac{n-j-k+q}{n-j}\right)^{\frac{1}{2}}|D_j^{(q)}\rangle|0\rangle|D_{n-j-1}^{(k-q)}\rangle+\\
\nonumber
& & \sum_{q^{\prime}=0}^{j}\left[\frac{(n-j)!k!(n-k)!}{n!(k-q^{\prime})!(n-k-j+q^{\prime})!}\left.j\choose q^{\prime} \right.\right]^{\frac{1}{2}}\\
\nonumber
& &\left(\frac{k-q^{\prime}}{n-j}\right)^{\frac{1}{2}}|D_j^{(q^{\prime})}\rangle|1\rangle|D_{n-j-1}^{(k-q^{\prime}-1)}\rangle.
\end{eqnarray}
We may rearrange the previous equation to get:
\begin{eqnarray}
\nonumber
|D_n^{(k)}\rangle & = & \sum_{q=0}^{j}\left[\frac{(n-j-1)!k!(n-k)!}{n!(k-q)!(n-k-j+q-1)!}\left.j\choose q\right.\right]^{\frac{1}{2}}\\
\nonumber
& & |D_j^{(q)}\rangle|0\rangle|D_{n-j-1}^{(k-q)}\rangle+\\
\nonumber
& & \sum_{q^{\prime}=0}^{j}\left[\frac{(n-j-1)!k!(n-k)!}{n!(k-q^{\prime}-1)!(n-k-j+q^{\prime})!}\left.j\choose q^{\prime} \right.\right]^{\frac{1}{2}}\\
& &|D_j^{(q^{\prime})}\rangle|1\rangle|D_{n-j-1}^{(k-q^{\prime}-1)}\rangle.
\end{eqnarray}
At this point it is convenient to write $q^{\prime}=q-1$ in the second sum in the previous equation. We get:%
\begin{eqnarray}
\nonumber
|D_n^{(k)}\rangle & = & \sum_{q=0}^{j}\left[\frac{(n-j-1)!k!(n-k)!}{n!(k-q)!(n-k-j-1+q)!}\left.j+1\choose q\right.\right]^{\frac{1}{2}}\\
\nonumber
& & \left(\frac{j+1-q}{j+1}\right)^{\frac{1}{2}}|D_j^{(q)}\rangle|0\rangle|D_{n-j-1}^{(k-q)}\rangle +\\
\nonumber
& & \sum_{q=1}^{j+1}\left[\frac{(n-j-1)!k!(n-k)!}{n!(k-q)!(n-k-j-1+q)!}\left.j+1 \choose q\right.\right]^{\frac{1}{2}}\\
& &\left(\frac{q}{j+1}\right)^{\frac{1}{2}}|D_j^{(q-1)}\rangle|1\rangle|D_{n-j-1}^{(k-q)}\rangle.
\end{eqnarray}
Now we define $j^{\prime}$ such that $j^{\prime}=j-1$, which leads to:
\begin{eqnarray}
\nonumber
|D_{n}^{(k)}\rangle & = & \sum_{q=0}^{j^\prime-1}\left[\frac{(n-j^\prime)!k!(n-k)!}{n!(k-q)!(n-k-j^\prime + q)!}\left.j^\prime \choose q\right.\right]^{\frac{1}{2}}\\
\nonumber
& &\left(\frac{j^\prime -q}{j^\prime}\right)^{\frac{1}{2}}|D_{j^\prime -1}^{(q)}\rangle|0\rangle|D_{n-j^{\prime}}^{(k-q)}\rangle+\\
\nonumber
& & \sum_{q=1}^{j^\prime}\left[\frac{(n-j^\prime)!k!(n-k)!}{n!(k-q)!(n-k-j^\prime+q)!}\left.j^\prime \choose q\right.\right]^{\frac{1}{2}}\\
& &\left(\frac{q}{j^\prime}\right)^{\frac{1}{2}}|D_{j^\prime -1}^{(q-1)}\rangle|1\rangle|D_{n-j^\prime}^{(k-q)}\rangle.
\end{eqnarray}
Using again relation (\ref{first}), Eq. (\ref{gen}) is recovered, thus completing the proof.
 \bibliography{bib}

\begin{thebibliography}{26}
\bibitem [{com()}]{comm}%
  \BibitemOpen
  \href@noop {} {}\bibinfo {note} {We could have defined the more general state
  $$|\phi_n^{(k)}\rangle=(1-a^2)^{\frac{1}{2}}|D_n^{(k)}\rangle+(a/\sqrt{\ell})\sum_{q=1}^{\ell}\hat
  P_q^{(n)}|\underbrace{0...0}_{n-k}\underbrace{1...1}_{k}\rangle.$$ However,
  for an increasing number of terms the level of asymmetry diminishes as, for
  $\ell=\left. n\choose k\right.$ we would get
  $|\phi_n^{(k)}\rangle=|D_n^{(k)}\rangle$. So the state used in the body of
  the paper, with $\ell=1$ is the most demanding for the witness.}\BibitemShut
  {Stop}%
\end{thebibliography}
\end{document}